\documentclass[a4paper,fleqn,usenatbib]{mnras}

\usepackage{newtxtext,newtxmath}
\usepackage[T1]{fontenc}
\usepackage{ae,aecompl}
\usepackage{graphicx}	
\usepackage{amsmath}	
\usepackage{amssymb}	
\usepackage{scalefnt}
\usepackage{threeparttable, tablefootnote}
\usepackage{wrapfig}	
\usepackage[textwidth=1.5cm,shadow]{todonotes}	
\usepackage{color, soul}	
\usepackage{verbatim}	

\def\apropto{%
  \def\p{%
    \setbox0=\vbox{\hbox{$\propto$}}%
    \ht0=0.6ex \box0 }%
  \def\s{%
    \vbox{\hbox{$\sim$}}%
  }%
  \mathrel{\raisebox{0.7ex}{%
      \mbox{$\underset{\s}{\p}$}%
    }}%
}

\title[NGC 3115's SED]{The multiwavelength spectrum of NGC 3115: Hot accretion flow properties}

\author[I. Almeida et al.]{
Ivan Almeida$^{1}$\thanks{E-mail: ivan.almeida@usp.br}, 
Rodrigo Nemmen$^{1}$, 
Ka-Wah Wong$^{2,3}$,
Qingwen Wu$^{4}$ 
\newauthor
and Jimmy A. Irwin$^{5}$
\\
$^{1}$Universidade de S\~ao Paulo, Instituto de Astronomia, Geof\'{\i}sica e Ci\^encias Atmosf\'ericas, Departamento de Astronomia,\\ S\~ao Paulo, SP 05508-090, Brazil\\
$^{2}$Eureka Scientific, Inc., 2452 Delmer Street Suite 100, Oakland, CA 94602-3017, USA \\
$^{3}$Department of Physics and Astronomy, Minnesota State University, Mankato, MN 56001, USA \\
$^{4}$School of Physics, Huazhong University of Science and Technology, Wuhan 430074, China \\
$^{5}$Department of Physics and Astronomy, University of Alabama, Box 870324, Tuscaloosa, AL 35487, USA
}

\date{Accepted 2018 January 11. Received 2018 January 9; in original form 2017 September 27}

\pubyear{2017}

\begin{document}
\label{firstpage}
\pagerange{\pageref{firstpage}--\pageref{lastpage}}
\maketitle

\begin{abstract}
NGC 3115 is the nearest galaxy hosting a billion solar mass black hole and is also a low-luminosity active galactic nucleus (LLAGN). X-ray observations of this LLAGN are able to spatially resolve the hot gas within the sphere of gravitational influence of the supermassive black hole. These observations make NGC 3115 an important testbed for black hole accretion theory in galactic nuclei since they constrain the outer boundary conditions of the hot accretion flow. We present a compilation of the multiwavelength spectral energy distribution (SED) of the nucleus of NGC 3115 from radio to X-rays. We report the results from modeling the observed SED with radiatively inefficient accretion flow (RIAF) models. The radio emission can be well-explained by synchrotron emission from the RIAF without the need for contribution from a relativistic jet. We obtain a tight constraint on the RIAF density profile, $\rho(r) \propto r^{-0.73 _{-0.02} ^{+0.01}}$, implying that mass-loss through subrelativistic outflows from the RIAF is significant. The lower frequency radio observation requires the synchrotron emission from a nonthermal electron population in the RIAF, similarly to Sgr A*. 
\end{abstract}


\section{Introduction}

NGC 3115 is a nearby lenticular galaxy \citep{Menezes2014} (distance of 9.7 Mpc) hosting the nearest $>10^9 M\odot$ black hole \citep{Kormendy1996,Emsellem1999}. Its central supermassive black hole (SMBH) is accreting at low rates with a Bondi accretion rate of $\dot{M}_{\rm B} = 2 \times 10^{-4} \dot{M}_{\rm Edd}$ \citep{Wong2014} where $\dot{M}_{\rm Edd}=10 L_{\rm Edd}/c^2$ is the Eddington mass accretion rate and hosts a low-luminosity AGN (LLAGN). As such, the SMBH in NGC 3115 is thought to be accreting in the RIAF mode (for a comprehensive review see \citealt{Yuan2014}). 

Thanks to the Megasecond \emph{Chandra} X-ray Visionary Project observation of NGC 3115's SMBH (PI: Irwin), \cite{Wong2011} placed the first direct observational constraints on the spatially and spectroscopically resolved structures of the X-ray emitting gas inside the Bondi radius ($R_B$) of a black hole. \cite{Wong2011,Wong2014} measured the temperature and density profiles of the hot gas from a fraction out to tens of the Bondi radius ($R_B = 2''.4 - 4''.8 = 112-224 \ {\rm pc}$). The density profile is broadly consistent with $\rho \propto r^{-1}$ similarly to the case of Sgr A* and M87 for which emission within $R_B$ can also be resolved with \emph{Chandra} \citep{Wang2013,Russell2015}. The observed density profile is in stark contrast with the $\rho \propto r^{-3/2}$ profile expected from Bondi accretion and simple RIAF models \citep{Narayan1994}. 

These \emph{Chandra} observations give us an unprecedented level of details on the gas surrounding the SMBH in NGC 3115, providing us with an unique characterization of the outer boundary conditions of an underfed SMBH in a nearby galaxy. Therefore, NGC 3115 is one of the best testbeds for the theory of low-density plasmas and hot accretion in the vicinity of compact objects. 

The goal of this paper is to compile the broadband, multiwavelength spectral energy distribution (SED) of the accreting SMBH in the nucleus of NGC 3115, and model the observed SED with radiative RIAF models--as appropriate for LLAGNs--assessing along the way whether theory reproduces observations. 

The structure of this paper is as follows. In Section \ref{sec:obs}, we describe the SED data compiled from previous observations. Section \ref{sec:model} describes the physical model that we adopted in order to interpret and fit the SEDs and the modeling results. In Section \ref{sec:disc} we contextualize our results given what is currently known from the observational and theoretical side of LLAGNs and black hole accretion theory. We conclude by presenting a summary of our results in Section \ref{sec:conc}.

\section{Observations}	\label{sec:obs}

We compiled the multiwavelength observations of NGC 3115 previously available in the literature, in order to gather the SED and proceed with the modeling. The data we were able to find cover from radio to X-rays energies; the observations were obtained with the Very Large Array (VLA), \textit{Spitzer}, GALEX, and \textit{Chandra}. All the SED data points are presented in Table \ref{tab:data} and displayed in Figure \ref{th3115}.We adopted the same distance for computing all luminosities.

\begin{table*}
\centering

\caption{SED data}
\label{data}
\begin{threeparttable}
\begin{tabular}{@{} l *4c @{}}
\multicolumn{1}{c}{$\nu$ (Hz)} & $\nu L_\nu$ ($erg\cdot s^{-1}$)& $\Delta \nu L_\nu$ ($erg\cdot s^{-1}$)  & Resolution \\ \hline
1.40E+09 \tnote{a}      & 6.92E+34   & 3.45E+34  &   5.9'' \\ 
5.00E+09 \tnote{b}      & < 1.86E+35     &           &    5'' \\ 
8.50E+09 \tnote{a}      & 3.10E+35 &  1.33E+35  &    0.17'' \\  
5.55E+14 \tnote{c}      & < 1.80E+40     &   &$\sim 0.1''$   \\ 
6.82E+14 \tnote{d}      & < 3.66E+38          &      &     $\sim 8''$   \\ 
4.84E+17 \tnote{d}      & < 5.51E+38           &     &    25''    \\ 
(1.2-14.5)E+17 \tnote{e} & < 4.4E+37 &  & 0.5'' \\
\hline
\end{tabular}
\begin{tablenotes}\footnotesize
\item References: 
\item [a] \cite{Wrobel2012}
\item [b] \cite{Fabbiano1989} 
\item [c] \cite{Lauer2005}  
\item [d] \cite{Wu2005}  
\item [e] \cite{Wong2014}
\end{tablenotes}
\end{threeparttable}
\label{tab:data}
\end{table*}

We treat the two observations that comprise the optical-UV portion of the SED as upper limits, since there is not enough information to infer the true emission of the LLAGN--our primary interest in this work--and thus these data points may potentially include considerable contamination by host galaxy emission \citep{Lauer2005}. 

\section{SED modeling}	\label{sec:model}

Given that NGC 3115 hosts a sub-Eddington LLAGN, presumably accreting in the RIAF mode, we use a semi-analytical approach to treat the radiation from this system in which the accretion flow is considered stationary assuming a $\alpha$-viscosity and a pseudo-Newtonian gravity, and the radiative transfer is treated in considerable detail, taking into account synchrotron, inverse Compton scattering and bremsstrahlung processes (e.g. \citealt{Yu2011, Nemmen2006, Nemmen2014}).

We do not consider the contribution to the emission by an optically thick, geometrically thin accretion disk for two reasons. Firstly, at the low accretion rates in NGC 3115 we do not expect a coherent thin disk \citep{Yuan2014}; secondly, there are not enough observational constraints in the SED in the near-IR and optical in order to meaningfully constrain the presence of the thin disk. Furthermore, given that NGC 3115 displays only compact radio emission and shows no evidence for the presence of extended jets we do not incorporate the contribution of a relativistic jet.

Our model for the RIAF emission follows \cite{Nemmen2014} (cf. also e.g. \citealt{Yu2011}). We now describe the main parameters of this model.
RIAFs are characterized by the presence of outflows or winds, which prevent a considerable fraction of the gas that is available at large radii from being accreted onto the black hole (see \citealt{Yuan2014} for a review on recent analytical and computational advances). In order to take this mass-loss into account, we introduce the parameter $s$ to describe the radial variation of the accretion rate as $\dot{M}(R) = \dot{M}_{\rm o} \left( R/R_{\rm o} \right)^{s}$ (or $\rho(R) \propto R^{-3/2+s}$) where $\dot{M}_{\rm o}$ is the rate measured at the outer radius $R_{\rm o}$ of the RIAF \citep{Blandford1999}. 

The other parameters that describe the RIAF solution are the black hole mass $M$; the viscosity parameter $\alpha$; the modified plasma $\beta$ parameter, defined as the ratio between the gas and total pressures, $\beta=P_g/P_{\rm tot}$; the fraction of energy dissipated via turbulence that directly heats electrons $\delta$; and the adiabatic index $\gamma$.

Following \cite{Nemmen2014}, in our calculations we adopt the typical choice of parameters $\alpha=0.3$, $\beta=0.9$, $\gamma=1.5$ and $R_o = 10^4 R_S$. Traditional RIAF models adopted $\delta$ to be small ($\delta \lesssim 0.01$; e.g., \citealt{Yuan2014}). 

Given considerable observational efforts to constrain the properties of the accreting black hole in NGC 3115, important empirical priors are available for our modeling leaving very little freedom in the SED model. We adopt the black hole mass $M=2 \times 10^9 M_\odot$ as estimated by \cite{Kormendy1996}. Using $M=10^9 M_\odot$ gives essentially the same results. We fix the outer accretion rate as $\dot{M}_o = \alpha \dot{M}_B = 6 \times 10^{-5} \dot{M}_{Edd}$ where we take into account the appropriate RIAF outer boundary conditions \citep{Narayan1995} and the Bondi rate constrained by \cite{Wong2014}. 

We end up with only two free parameters in the modeling: The $s$ parameter that sets the RIAF density profile and $\delta$. We only allow $s$ to vary within the limits obtained from \emph{Chandra} X-ray constraints on the density profile within 3'' (140 pc) by \cite{Wong2014}, roughly $s \approx 0.6-1.26$. Given the theoretical uncertainty related to the value of $\delta$, we allow it to vary over the range $0.01 \leq \delta \leq 0.5$ \citep{Sharma2007,Howes2010}.

In this work we adopted a iterative procedure where we changed individually the parameters, keeping the others fixed, until we find the set of values that best reproduces visually the observations \citep{Nemmen2014}. With only two free parameters, this task is made immensely more convenient. 

\begin{figure}
\centering      
\includegraphics[width=\linewidth,trim=0 0 0 0,clip=true]{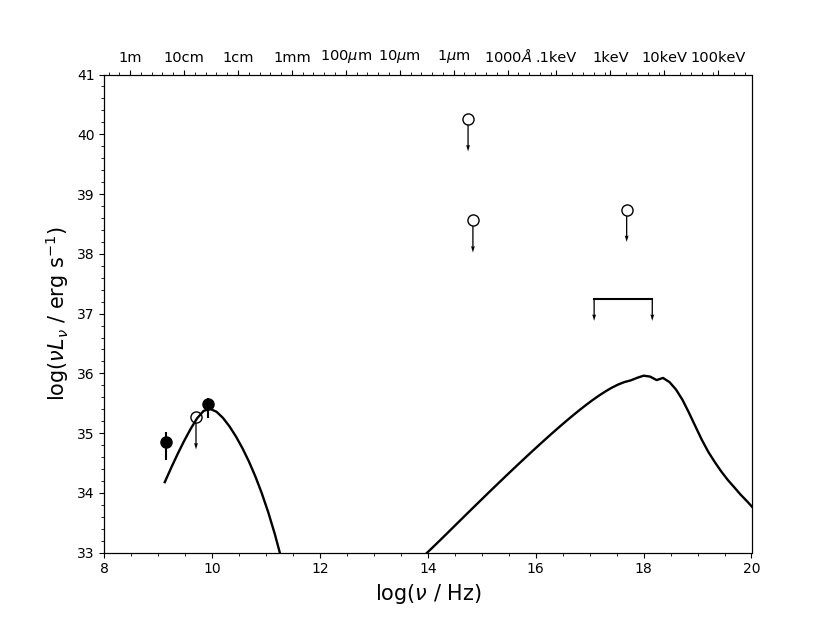}
\caption{SED of NGC 3115 with table \ref{tab:data} data. The plot is the best thermal obtained fit, with s = 0.77 and $\delta$ = 0.095.}
\label{th3115}
\end{figure}

In figure \ref{th3115} we show the observations and the best model fit. The solid line is the predicted spectrum of the accretion flow for NGC 3115 with s = 0.77 and $\delta$ = 0.095. The model is consistent with all observations except the lower frequency radio point at 1.4 GHz, which is underpredicted. We estimated a rough uncertainty of 0.02 in the value of $s$ in order to fit the radio data. Keeping all parameters fixed to the above values and varying only $\delta$, we estimate that a rough uncertainty on 0.01 in $\delta$ is consistent with the observations.  

It is known that the compact radio emission in LLAGNs is underpredicted by models which only account for synchrotron emission from a thermal distribution of electrons in the RIAF. In the case of radio-loud LLAGNs, the excess radio emission is naturally explained by the relativistic jet (e.g., \citealt{Wu2007,Yu2011,Nemmen2014}). \cite{Liu2013} modeled a sample of LLAGNs where they excluded the sources which display evidence of resolved radio jets; they found that the compact radio emission for this sample is well-explained by an alternative model where the synchrotron emission includes the contribution of a nonthermal electron distribution in the RIAF--in addition to a thermal component, similar to Sgr A* \citep{Yuan2003}. In these models, the origin of the nonthermal electron component is due to a non-specified particle acceleration process such as Fermi first order process (e.g. \citealt{Sironi2015}).

We incorporate the nonthermal electron distribution following the approach of \cite{Liu2013}.
The thermal electrons are assumed to follow a relativistic Maxwell-Boltzmann distribution and the nonthermal electrons follow a power-law tail $n_{\rm e}\propto \gamma^{-p}$ where $n_{\rm e}$ is the number density of nonthermal electrons and $\gamma$ is the electron Lorentz factor. Assuming the energy in nonthermal electrons is a fraction $\eta$ of the energy in thermal electrons, the number density of nonthermal electrons can be derived. The synchrotron emission from thermal electrons and nonthermal electrons  can be calculated with the above thermal and nonthermal electron distributions.

\begin{figure}
\centering      
\includegraphics[width=\linewidth,trim=0 0 0 0,clip=true]{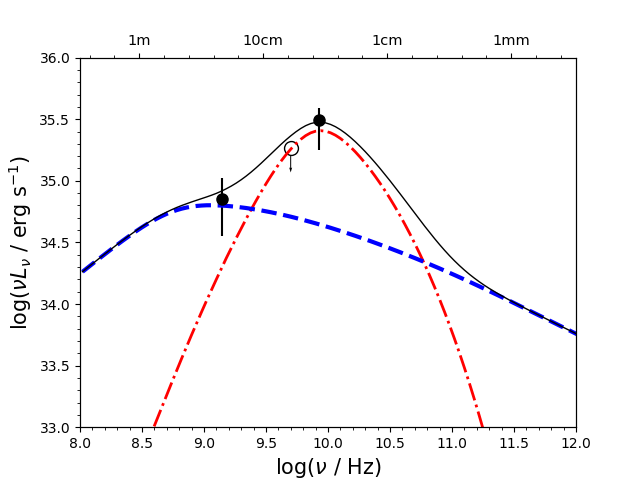}
\caption{Similar to Figure 1, but with emission from non-thermal electrons included in the model. The dashed line represents emission from the non-thermal electrons and the dash-dotted line represents the thermal emission. The solid line is the sum of the two components.}
\label{nth3115}
\end{figure}

The RIAF SED model incorporating nonthermal electrons successfully explains the 1.4 GHz data point, as can be seen in Figure \ref{nth3115}. This figure shows the best-fit nonthermal model where we fixed all parameters shared with the thermal-only model. The resulting values for the free parameters in this case are $s = 0.76$, $\delta = 0.1$ and $\eta$ = 3\%, where we adopted a typical value of $p = 3$ for power-law index for the non-thermal electron distribution. 

\section{Discussion}	\label{sec:disc}

In our modeling of the SED, we obtained a rough constraint on the fraction of the turbulent energy that heats the electrons as $\delta = 0.095 \pm 0.01$; $\delta$-values outside this range--for a fixed value of $s$--considerably underpredict or overpredict the radio emission. The resulting value of $0.095$ is comfortably within the range of values $\sim 0.01-0.5$ predicted by more sophisticated models for the electron heating  in RIAFs (\citealt{Bisnovatyi-Kogan1997, Quataert1999turb, Sharma2007,Howes2010}; cf. \citealt{Xie2012} for an overview).

\emph{Chandra} X-ray observations of the hot diffuse gas on scales of $\sim 100 pc$ (2'') placed constraints on the density profile of $s \approx 0.72-1.11$ ($\rho(r) \propto r^{-(0.39-0.78)}$) \citep{Wong2014}. The results of our modeling place much tighter constraints on the inner density profile for the accretion flow in NGC 3115. We found the resulting range $s=0.76-0.79$; the higher $s$-values are from the SED models including only thermal electrons whereas the lower-end $s$-values come from the models with nonthermal electron contribution. Apart from Sgr A* \citep{Yuan2003,Yuan2006}, this is the best constraint for the density profile--and correspondingly mass-loss ratio--in a LLAGN. 

The above constraint on $s$ should provide valuable input to studies of the dynamics of RIAFs. For instance, they suggest that the amount of mass-loss through winds in RIAFs is not as extreme as proposed by some authors. For example, \cite{Begelman2012} revised the adiabatic inflow-outflow solution (ADIOS) model and proposed that $s$ should be 1. This is disfavored by our results; on the contrary, this seems to give support to the numerical simulations of \cite{Yuan2012,Yuan2012b}. \cite{Yuan2012} performed 2D numerical hydrodynamical simulations of RIAFs with a large dynamical range; they found $s \approx 0.5-0.7$. Our results are more consistent with Yuan's work.

Related to the nonthermal electron RIAF model, the parameters $\eta$ and $p$ are degenerated in reproducing the low-frequency radio spectrum (e.g., \citealt{Ozel2000}), where $\eta$ will be lower for smaller $p$. For a typical value $p=3$ we obtain $\eta \sim 3\%$ in NGC 3115, which is quite consistent with the constraints obtained from nearby LLAGNs and Sgr A* (e.g., $\eta \sim 1\%$; \citealt{Yuan2003, Liu2013}).

It is worth comparing our results with those of \cite{Shcherbakov2014} who also modeled the gas accretion in NGC 3115. The main difference in their approach is that they take into account gas injection by stars in the nuclear star cluster and the effect of energy injection by supernova explosions, electron heat conduction and Coulomb collisions--among other effects--and do not include viscosity and rotation. They compared their model with the spatially resolved X-ray observations and from their fits obtained an upper limit to the mass accretion rate onto the SMBH of $\dot{M} = 2 \times 10^{-3} \ M_\odot \ {\rm yr}^{-1}$. From our best-fit RIAF model, $\dot{M}(R_o) = 3 \times 10^{-3} \ M_\odot \ {\rm yr}^{-1}$ at the RIAF outer radius $R_o$. Given the considerable amount of mass-loss in the accretion flow, $\dot{M}(3 R_S) = 5 \times 10^{-6} \ M_\odot \ {\rm yr}^{-1}$ at the innermost stable circular orbit for a Schwarzschild black hole; this is consistent with the \cite{Shcherbakov2014} accretion rate upper limit. Furthermore, they estimate a shallow density profile $\rho \apropto r^{-1}$ over a large dynamic range. Therefore, their estimated radial dependency of the density is roughly in agreement with our results.

One model that remains to be better explored in the future for NGC 3115 is the chaotic cold accretion (CCA; \citealt{Gaspari2013,Gaspari2017}) flow. The CCA model takes into account cooling effects and associated thermal instabilities as the gas flows from the larger galactic scales down to the SMBH, which could lead to the condensation of cold clouds. Interestingly enough, the CCA model predicts $\rho(r) \propto r^{-(0.7-1)}$ (Gaspari, private communication) which is in agreement with our results. The cold and chaotic mode of accretion in the CCA model boosts the SMBH accretion rate to levels of up to $100 \dot{M}_B$; this is in stark contrast with our work, where we fix the outer accretion rate to $\dot{M}(R_o) \sim 0.1 \dot{M}_B$ according to RIAF theory expectations (\textsection \ref{sec:model}). It will be worth in the future to test the CCA model with e.g. Integral Field Unit observations \citep{Gaspari2017a} of NGC 3115.

We should now point out some of the limitations in our model. The dynamical structure of the flow is based on the solution to stationary, height-integrated, one-dimensional equations describing a RIAF in a pseudo-Newtonian potential \citep{Narayan1995,Yuan2008} adopting an $\alpha$-prescription for the viscosity \citep{Shakura1973}. The radiative transfer is carried out on top of this simplified flow structure with considerable detail \citep{Yuan2005,Nemmen2014}. 
Nowadays, there is considerable interest in treating in more details the dynamics and radiation spectra of RIAFs using multidimensional, general relativistic radiative magnetohydrodynamics (GRRMHD) numerical simulations (e.g. \citealt{Chan2015,Ryan2015,Riordan2016,Ryan2017}). We should stress that 1D stationary RIAF spectral models are still very useful even though sophisticated 3D GRRMHD simulations are available. The 1D models are much computationally faster for generating SED models to compare with observations, whereas in the case of 3D simulations this is not usually practical. Furthermore, 1D models can provide useful insights to understand the physics, something which is harder to get from 3D GRRMHD simulations.  

One may wonder how the resulting SED and radiative efficiency from our simplified models compare with those of more advanced, numerical simulations. Recently, \cite{Ryan2017} performed a GRRMHD simulation of an RIAF including a detailed treatment of the electron thermodynamics and evolving a two-temperature plasma and a frequency-dependent radiative transport. This simulation is able to predict multiwavelength SEDs and the behavior of the radiative efficiency $\epsilon \equiv L/(\dot{M} c^2)$ as a function of $\dot{M}$. We estimate the accretion rate at the black hole from our SED models as $\dot{M} = (R_S/R_o)^s \approx 10^{-7} \dot{M}_{\rm Edd}$; the corresponding radiative efficiency in our model is $\epsilon \approx 10^{-4}$ \citep{Xie2012}. Satisfyingly, \cite{Ryan2017} computed $\epsilon = 2.6 \times 10^{-4}$ from their GRRMHD model for a comparable accretion rate. This is a neat consistency check of our calculations.

\section{Conclusions}	\label{sec:conc}

NGC 3115 is the only galactic nucleus besides Sgr A* and M87 for which the the sphere of gravitational influence of the black hole--the Bondi radius--can be spatially resolved in X-rays. Due to the well-determined Bondi accretion rate from \textit{Chandra} observations, there are only two free parameters in the SED modeling which are related to the density profile  and the electron heating rate in the accretion flow, i.e. the global dynamics and the electron thermodynamics. We have analyzed the broadband, multiwavelength SED of NGC 3115 in order to obtain information about the physics of the massive black hole in its low-luminosity AGN. Our main conclusions are the following:
\begin{itemize}
\item NGC 3115 does not require a relativistic jet in order to account for the radio emission: it is well-explained by synchrotron emission from the hot accretion flow.
\item The lower frequency (1.4 GHz) radio observation is underpredicted by a thermal synchrotron emission from the RIAF. This suggests that a small fraction of nonthermal electrons is required to account for this observation, similar to Sgr A*. 
\item We obtained an independent constraint $\rho \propto r^{-0.73 ^{+0.01} _{-0.02}}$ or $\dot{M} \propto r^{0.77 _{-0.01} ^{+0.02}}$ on the inner density profile. This result implies that mass-loss through subrelativistic outflows from the RIAF is significant in this accreting system. This is also consistent with previous \textit{Chandra} estimates as well as theoretical RIAF expectations about the role of winds.
\item We constrain the fraction of viscous energy that directly heats electrons as $\delta \approx 0.095$. This is consistent with theoretical studies of dissipation in collisionless plasmas which suggest that electrons receive a significant fraction of the viscously dissipated energy in the flow.
\item The radiative efficiency from our stationary, one-dimensional model, $L/(\dot{M} c^2) \approx 10^{-4}$, is very similar to the corresponding efficiency resulting from a recent general relativistic radiative MHD numerical simulation for a comparable accretion rate.
\end{itemize}

The fact that this source lacks an extended relativistic jet suggests that the central black hole is slowly spinning, otherwise a stronger jet would be produced via the Blandford-Znajek mechanism (e.g., \citealt{Blandford1977,Sasha2011}). Since the size and shape of a rapidly spinning SMBH is markedly different from that of a Schwarzschild hole, the slowly-spinning hypothesis could in principle be tested with Event Horizon Telescope (EHT) observations \citep{Doeleman2012} resolving the shadow cast by the SMBH in NGC 3115. The predicted size of the event horizon shadow for a Schwarzschild black hole is $2 \sqrt{27} GM/c^2 \approx 10 \mu$as \citep{Bardeen1973,Luminet1979} which could be resolvable with the EHT at mm-wavelengths \citep{Falcke2013}. However, NGC 3115's LLAGN flux is only $10^{-7}$ of Sgr A*'s flux; it will be very hard to even detect any radio emission at all--and much less a SMBH shadow--with the EHT sensitivity for this faint LLAGN.

There are only a few high-angular-resolution observations of the NGC 3115's nucleus; most of the measurements were obtained with low angular resolution and are contaminated by galactic emission (e.g. stellar populations, dust) as a consequence. More high-angular-resolution observations for this source in the mm to X-rays range will be instrumental in order to better test the physics of accretion flows and collisionless plasmas in LLAGNs.

\section*{Acknowledgements}
We acknowledge the help of Roberto Menezes, Rog\'erio Riffel and Rogemar Riffel in reducing a Gemini spectrum of NGC 3115, the continuum of which unfortunately was dominated by stellar emission and contained no information on the LLAGN. We acknowledge useful discussions with Jo\~ao Steiner, Alexander Tchekhovskoy and Massimo Gaspari. This work was supported by FAPESP (Funda\c{c}\~ao de Amparo \`a Pesquisa do Estado de S\~ao Paulo) under grants 2015/26831-1 and 2016/24857-6.

\bibliographystyle{mnras}
\bibliography{refs2}

\begin{thebibliography}{}
\makeatletter
\relax
\def\mn@urlcharsother{\let\do\@makeother \do\$\do\&\do\#\do\^\do\_\do\%\do\~}
\def\mn@doi{\begingroup\mn@urlcharsother \@ifnextchar [ {\mn@doi@}
  {\mn@doi@[]}}
\def\mn@doi@[#1]#2{\def\@tempa{#1}\ifx\@tempa\@empty \href
  {http://dx.doi.org/#2} {doi:#2}\else \href {http://dx.doi.org/#2} {#1}\fi
  \endgroup}
\def\mn@eprint#1#2{\mn@eprint@#1:#2::\@nil}
\def\mn@eprint@arXiv#1{\href {http://arxiv.org/abs/#1} {{\tt arXiv:#1}}}
\def\mn@eprint@dblp#1{\href {http://dblp.uni-trier.de/rec/bibtex/#1.xml}
  {dblp:#1}}
\def\mn@eprint@#1:#2:#3:#4\@nil{\def\@tempa {#1}\def\@tempb {#2}\def\@tempc
  {#3}\ifx \@tempc \@empty \let \@tempc \@tempb \let \@tempb \@tempa \fi \ifx
  \@tempb \@empty \def\@tempb {arXiv}\fi \@ifundefined
  {mn@eprint@\@tempb}{\@tempb:\@tempc}{\expandafter \expandafter \csname
  mn@eprint@\@tempb\endcsname \expandafter{\@tempc}}}

\bibitem[\protect\citeauthoryear{{Bardeen}}{{Bardeen}}{1973}]{Bardeen1973}
{Bardeen} J.~M.,  1973, in {Dewitt} C.,  {Dewitt} B.~S.,  eds, Black Holes (Les
  Astres Occlus). pp 215--239

\bibitem[\protect\citeauthoryear{{Begelman}}{{Begelman}}{2012}]{Begelman2012}
{Begelman} M.~C.,  2012, \mn@doi [\mnras] {10.1111/j.1365-2966.2011.20071.x},
  \href {http://adsabs.harvard.edu/abs/2012MNRAS.420.2912B} {420, 2912}

\bibitem[\protect\citeauthoryear{{Bisnovatyi-Kogan} \&
  {Lovelace}}{{Bisnovatyi-Kogan} \& {Lovelace}}{1997}]{Bisnovatyi-Kogan1997}
{Bisnovatyi-Kogan} G.~S.,  {Lovelace} R.~V.~E.,  1997, \mn@doi [\apjl]
  {10.1086/310826}, \href {http://adsabs.harvard.edu/abs/1997ApJ...486L..43B}
  {486, L43}

\bibitem[\protect\citeauthoryear{{Blandford} \& {Begelman}}{{Blandford} \&
  {Begelman}}{1999}]{Blandford1999}
{Blandford} R.~D.,  {Begelman} M.~C.,  1999, \mn@doi [\mnras]
  {10.1046/j.1365-8711.1999.02358.x}, \href
  {http://adsabs.harvard.edu/abs/1999MNRAS.303L...1B} {303, L1}

\bibitem[\protect\citeauthoryear{{Blandford} \& {Znajek}}{{Blandford} \&
  {Znajek}}{1977}]{Blandford1977}
{Blandford} R.~D.,  {Znajek} R.~L.,  1977, \mnras, \href
  {http://adsabs.harvard.edu/abs/1977MNRAS.179..433B} {179, 433}

\bibitem[\protect\citeauthoryear{{Chan}, {Psaltis}, {{\"O}zel}, {Medeiros},
  {Marrone}, {S{\c a}dowski}  \& {Narayan}}{{Chan} et~al.}{2015}]{Chan2015}
{Chan} C.-k.,  {Psaltis} D.,  {{\"O}zel} F.,  {Medeiros} L.,  {Marrone} D.,
  {S{\c a}dowski} A.,   {Narayan} R.,  2015, \mn@doi [\apj]
  {10.1088/0004-637X/812/2/103}, \href
  {http://adsabs.harvard.edu/abs/2015ApJ...812..103C} {812, 103}

\bibitem[\protect\citeauthoryear{{Doeleman} et~al.,}{{Doeleman}
  et~al.}{2012}]{Doeleman2012}
{Doeleman} S.~S.,  et~al., 2012, \mn@doi [Science] {10.1126/science.1224768},
  \href {http://adsabs.harvard.edu/abs/2012Sci...338..355D} {338, 355}

\bibitem[\protect\citeauthoryear{{Emsellem}, {Dejonghe}  \& {Bacon}}{{Emsellem}
  et~al.}{1999}]{Emsellem1999}
{Emsellem} E.,  {Dejonghe} H.,   {Bacon} R.,  1999, \mn@doi [\mnras]
  {10.1046/j.1365-8711.1999.02210.x}, \href
  {http://adsabs.harvard.edu/abs/1999MNRAS.303..495E} {303, 495}

\bibitem[\protect\citeauthoryear{Fabbiano, Gioia  \& Trinchieri}{Fabbiano
  et~al.}{1989}]{Fabbiano1989}
Fabbiano G.,  Gioia I.,   Trinchieri G.,  1989, \apj, 347, 127

\bibitem[\protect\citeauthoryear{{Falcke} \& {Markoff}}{{Falcke} \&
  {Markoff}}{2013}]{Falcke2013}
{Falcke} H.,  {Markoff} S.~B.,  2013, \mn@doi [Classical and Quantum Gravity]
  {10.1088/0264-9381/30/24/244003}, \href
  {http://adsabs.harvard.edu/abs/2013CQGra..30x4003F} {30, 244003}

\bibitem[\protect\citeauthoryear{{Gaspari}, {Ruszkowski}  \& {Oh}}{{Gaspari}
  et~al.}{2013}]{Gaspari2013}
{Gaspari} M.,  {Ruszkowski} M.,   {Oh} S.~P.,  2013, \mn@doi [\mnras]
  {10.1093/mnras/stt692}, \href
  {http://adsabs.harvard.edu/abs/2013MNRAS.432.3401G} {432, 3401}

\bibitem[\protect\citeauthoryear{{Gaspari} et~al.,}{{Gaspari}
  et~al.}{2017a}]{Gaspari2017a}
{Gaspari} M.,  et~al., 2017a, preprint, \href
  {http://adsabs.harvard.edu/abs/2017arXiv170906564G} {} (\mn@eprint {arXiv}
  {1709.06564})

\bibitem[\protect\citeauthoryear{{Gaspari}, {Temi}  \& {Brighenti}}{{Gaspari}
  et~al.}{2017b}]{Gaspari2017}
{Gaspari} M.,  {Temi} P.,   {Brighenti} F.,  2017b, \mn@doi [\mnras]
  {10.1093/mnras/stw3108}, \href
  {http://adsabs.harvard.edu/abs/2017MNRAS.466..677G} {466, 677}

\bibitem[\protect\citeauthoryear{{Howes}}{{Howes}}{2010}]{Howes2010}
{Howes} G.~G.,  2010, \mn@doi [\mnras] {10.1111/j.1745-3933.2010.00958.x},
  \href {http://adsabs.harvard.edu/abs/2010MNRAS.409L.104H} {409, L104}

\bibitem[\protect\citeauthoryear{{Kormendy} et~al.,}{{Kormendy}
  et~al.}{1996}]{Kormendy1996}
{Kormendy} J.,  et~al., 1996, \mn@doi [\apjl] {10.1086/309950}, \href
  {http://adsabs.harvard.edu/abs/1996ApJ...459L..57K} {459, L57}

\bibitem[\protect\citeauthoryear{Lauer et~al.,}{Lauer et~al.}{2005}]{Lauer2005}
Lauer T.~R.,  et~al., 2005, The Astronomical Journal, 129, 2138

\bibitem[\protect\citeauthoryear{{Liu} \& {Wu}}{{Liu} \& {Wu}}{2013}]{Liu2013}
{Liu} H.,  {Wu} Q.,  2013, \mn@doi [\apj] {10.1088/0004-637X/764/1/17}, \href
  {http://adsabs.harvard.edu/abs/2013ApJ...764...17L} {764, 17}

\bibitem[\protect\citeauthoryear{{Luminet}}{{Luminet}}{1979}]{Luminet1979}
{Luminet} J.-P.,  1979, \aap, \href
  {http://adsabs.harvard.edu/abs/1979A%26A....75..228L} {75, 228}

\bibitem[\protect\citeauthoryear{{Menezes}, {Steiner}  \& {Ricci}}{{Menezes}
  et~al.}{2014}]{Menezes2014}
{Menezes} R.~B.,  {Steiner} J.~E.,   {Ricci} T.~V.,  2014, \mn@doi [\apjl]
  {10.1088/2041-8205/796/1/L13}, \href
  {http://adsabs.harvard.edu/abs/2014ApJ...796L..13M} {796, L13}

\bibitem[\protect\citeauthoryear{{Narayan} \& {Yi}}{{Narayan} \&
  {Yi}}{1994}]{Narayan1994}
{Narayan} R.,  {Yi} I.,  1994, \mn@doi [\apjl] {10.1086/187381}, \href
  {http://adsabs.harvard.edu/abs/1994ApJ...428L..13N} {428, L13}

\bibitem[\protect\citeauthoryear{{Narayan} \& {Yi}}{{Narayan} \&
  {Yi}}{1995}]{Narayan1995}
{Narayan} R.,  {Yi} I.,  1995, \mn@doi [\apj] {10.1086/176343}, \href
  {http://adsabs.harvard.edu/abs/1995ApJ...452..710N} {452, 710}

\bibitem[\protect\citeauthoryear{{Nemmen}, {Storchi-Bergmann}, {Yuan},
  {Eracleous}, {Terashima}  \& {Wilson}}{{Nemmen} et~al.}{2006}]{Nemmen2006}
{Nemmen} R.~S.,  {Storchi-Bergmann} T.,  {Yuan} F.,  {Eracleous} M.,
  {Terashima} Y.,   {Wilson} A.~S.,  2006, \mn@doi [\apj] {10.1086/500571},
  \href {http://adsabs.harvard.edu/abs/2006ApJ...643..652N} {643, 652}

\bibitem[\protect\citeauthoryear{{Nemmen}, {Storchi-Bergmann}  \&
  {Eracleous}}{{Nemmen} et~al.}{2014}]{Nemmen2014}
{Nemmen} R.~S.,  {Storchi-Bergmann} T.,   {Eracleous} M.,  2014, \mn@doi
  [\mnras] {10.1093/mnras/stt2388}, \href
  {http://adsabs.harvard.edu/abs/2014MNRAS.438.2804N} {438, 2804}

\bibitem[\protect\citeauthoryear{O'Riordan, Pe'er  \& McKinney}{O'Riordan
  et~al.}{2016}]{Riordan2016}
O'Riordan M.,  Pe'er A.,   McKinney J.~C.,  2016, \apj, 819, 95

\bibitem[\protect\citeauthoryear{{{\"O}zel}, {Psaltis}  \&
  {Narayan}}{{{\"O}zel} et~al.}{2000}]{Ozel2000}
{{\"O}zel} F.,  {Psaltis} D.,   {Narayan} R.,  2000, \mn@doi [\apj]
  {10.1086/309396}, \href {http://adsabs.harvard.edu/abs/2000ApJ...541..234O}
  {541, 234}

\bibitem[\protect\citeauthoryear{{Quataert} \& {Gruzinov}}{{Quataert} \&
  {Gruzinov}}{1999}]{Quataert1999turb}
{Quataert} E.,  {Gruzinov} A.,  1999, \mn@doi [\apj] {10.1086/307423}, \href
  {http://adsabs.harvard.edu/abs/1999ApJ...520..248Q} {520, 248}

\bibitem[\protect\citeauthoryear{{Russell}, {Fabian}, {McNamara}  \&
  {Broderick}}{{Russell} et~al.}{2015}]{Russell2015}
{Russell} H.~R.,  {Fabian} A.~C.,  {McNamara} B.~R.,   {Broderick} A.~E.,
  2015, \mn@doi [\mnras] {10.1093/mnras/stv954}, \href
  {http://adsabs.harvard.edu/abs/2015MNRAS.451..588R} {451, 588}

\bibitem[\protect\citeauthoryear{{Ryan}, {Dolence}  \& {Gammie}}{{Ryan}
  et~al.}{2015}]{Ryan2015}
{Ryan} B.~R.,  {Dolence} J.~C.,   {Gammie} C.~F.,  2015, \mn@doi [\apj]
  {10.1088/0004-637X/807/1/31}, \href
  {http://adsabs.harvard.edu/abs/2015ApJ...807...31R} {807, 31}

\bibitem[\protect\citeauthoryear{{Ryan}, {Ressler}, {Dolence}, {Tchekhovskoy},
  {Gammie}  \& {Quataert}}{{Ryan} et~al.}{2017}]{Ryan2017}
{Ryan} B.~R.,  {Ressler} S.~M.,  {Dolence} J.~C.,  {Tchekhovskoy} A.,  {Gammie}
  C.,   {Quataert} E.,  2017, \mn@doi [\apjl] {10.3847/2041-8213/aa8034}, \href
  {http://adsabs.harvard.edu/abs/2017ApJ...844L..24R} {844, L24}

\bibitem[\protect\citeauthoryear{{Shakura} \& {Sunyaev}}{{Shakura} \&
  {Sunyaev}}{1973}]{Shakura1973}
{Shakura} N.~I.,  {Sunyaev} R.~A.,  1973, \aap, \href
  {http://adsabs.harvard.edu/abs/1973A%26A....24..337S} {24, 337}

\bibitem[\protect\citeauthoryear{{Sharma}, {Quataert}, {Hammett}  \&
  {Stone}}{{Sharma} et~al.}{2007}]{Sharma2007}
{Sharma} P.,  {Quataert} E.,  {Hammett} G.~W.,   {Stone} J.~M.,  2007, \mn@doi
  [\apj] {10.1086/520800}, \href
  {http://adsabs.harvard.edu/abs/2007ApJ...667..714S} {667, 714}

\bibitem[\protect\citeauthoryear{{Shcherbakov}}{{Shcherbakov}}{2014}]{Shcherbakov2014}
{Shcherbakov} R.~V.,  2014, \mn@doi [\apj] {10.1088/0004-637X/783/1/31}, \href
  {http://adsabs.harvard.edu/abs/2014ApJ...783...31S} {783, 31}

\bibitem[\protect\citeauthoryear{{Sironi}, {Keshet}  \& {Lemoine}}{{Sironi}
  et~al.}{2015}]{Sironi2015}
{Sironi} L.,  {Keshet} U.,   {Lemoine} M.,  2015, \mn@doi [\ssr]
  {10.1007/s11214-015-0181-8}, \href
  {http://adsabs.harvard.edu/abs/2015SSRv..191..519S} {191, 519}

\bibitem[\protect\citeauthoryear{{Tchekhovskoy}, {Narayan}  \&
  {McKinney}}{{Tchekhovskoy} et~al.}{2011}]{Sasha2011}
{Tchekhovskoy} A.,  {Narayan} R.,   {McKinney} J.~C.,  2011, \mn@doi [\mnras]
  {10.1111/j.1745-3933.2011.01147.x}, \href
  {http://adsabs.harvard.edu/abs/2011MNRAS.418L..79T} {418, L79}

\bibitem[\protect\citeauthoryear{Wang et~al.,}{Wang et~al.}{2013}]{Wang2013}
Wang Q.~D.,  et~al., 2013, \mn@doi [Science] {10.1126/science.1240755}, 341,
  981

\bibitem[\protect\citeauthoryear{{Wong}, {Irwin}, {Yukita}, {Million},
  {Mathews}  \& {Bregman}}{{Wong} et~al.}{2011}]{Wong2011}
{Wong} K.-W.,  {Irwin} J.~A.,  {Yukita} M.,  {Million} E.~T.,  {Mathews} W.~G.,
    {Bregman} J.~N.,  2011, \mn@doi [\apjl] {10.1088/2041-8205/736/1/L23},
  \href {http://adsabs.harvard.edu/abs/2011ApJ...736L..23W} {736, L23}

\bibitem[\protect\citeauthoryear{{Wong}, {Irwin}, {Shcherbakov}, {Yukita},
  {Million}  \& {Bregman}}{{Wong} et~al.}{2014}]{Wong2014}
{Wong} K.-W.,  {Irwin} J.~A.,  {Shcherbakov} R.~V.,  {Yukita} M.,  {Million}
  E.~T.,   {Bregman} J.~N.,  2014, \mn@doi [\apj] {10.1088/0004-637X/780/1/9},
  \href {http://adsabs.harvard.edu/abs/2014ApJ...780....9W} {780, 9}

\bibitem[\protect\citeauthoryear{{Wrobel} \& {Nyland}}{{Wrobel} \&
  {Nyland}}{2012}]{Wrobel2012}
{Wrobel} J.~M.,  {Nyland} K.,  2012, \mn@doi [\aj]
  {10.1088/0004-6256/144/6/160}, \href
  {http://adsabs.harvard.edu/abs/2012AJ....144..160W} {144, 160}

\bibitem[\protect\citeauthoryear{{Wu} \& {Cao}}{{Wu} \& {Cao}}{2005}]{Wu2005}
{Wu} Q.,  {Cao} X.,  2005, \mn@doi [\apj] {10.1086/427428}, \href
  {http://adsabs.harvard.edu/abs/2005ApJ...621..130W} {621, 130}

\bibitem[\protect\citeauthoryear{{Wu}, {Yuan}  \& {Cao}}{{Wu}
  et~al.}{2007}]{Wu2007}
{Wu} Q.,  {Yuan} F.,   {Cao} X.,  2007, \mn@doi [\apj] {10.1086/521212}, \href
  {http://adsabs.harvard.edu/abs/2007ApJ...669...96W} {669, 96}

\bibitem[\protect\citeauthoryear{{Xie} \& {Yuan}}{{Xie} \&
  {Yuan}}{2012}]{Xie2012}
{Xie} F.-G.,  {Yuan} F.,  2012, \mn@doi [\mnras]
  {10.1111/j.1365-2966.2012.22030.x}, \href
  {http://adsabs.harvard.edu/abs/2012MNRAS.427.1580X} {427, 1580}

\bibitem[\protect\citeauthoryear{{Yu}, {Yuan}  \& {Ho}}{{Yu}
  et~al.}{2011}]{Yu2011}
{Yu} Z.,  {Yuan} F.,   {Ho} L.~C.,  2011, \mn@doi [\apj]
  {10.1088/0004-637X/726/2/87}, \href
  {http://adsabs.harvard.edu/abs/2011ApJ...726...87Y} {726, 87}

\bibitem[\protect\citeauthoryear{{Yuan} \& {Narayan}}{{Yuan} \&
  {Narayan}}{2014}]{Yuan2014}
{Yuan} F.,  {Narayan} R.,  2014, \mn@doi [\araa]
  {10.1146/annurev-astro-082812-141003}, \href
  {http://adsabs.harvard.edu/abs/2014ARA%26A..52..529Y} {52, 529}

\bibitem[\protect\citeauthoryear{{Yuan}, {Quataert}  \& {Narayan}}{{Yuan}
  et~al.}{2003}]{Yuan2003}
{Yuan} F.,  {Quataert} E.,   {Narayan} R.,  2003, \mn@doi [\apj]
  {10.1086/378716}, \href {http://adsabs.harvard.edu/abs/2003ApJ...598..301Y}
  {598, 301}

\bibitem[\protect\citeauthoryear{{Yuan}, {Cui}  \& {Narayan}}{{Yuan}
  et~al.}{2005}]{Yuan2005}
{Yuan} F.,  {Cui} W.,   {Narayan} R.,  2005, \mn@doi [\apj] {10.1086/427206},
  \href {http://adsabs.harvard.edu/abs/2005ApJ...620..905Y} {620, 905}

\bibitem[\protect\citeauthoryear{{Yuan}, {Shen}  \& {Huang}}{{Yuan}
  et~al.}{2006}]{Yuan2006}
{Yuan} F.,  {Shen} Z.-Q.,   {Huang} L.,  2006, \mn@doi [\apjl]
  {10.1086/504475}, \href {http://adsabs.harvard.edu/abs/2006ApJ...642L..45Y}
  {642, L45}

\bibitem[\protect\citeauthoryear{{Yuan}, {Ma}  \& {Narayan}}{{Yuan}
  et~al.}{2008}]{Yuan2008}
{Yuan} F.,  {Ma} R.,   {Narayan} R.,  2008, \mn@doi [\apj] {10.1086/587484},
  \href {http://adsabs.harvard.edu/abs/2008ApJ...679..984Y} {679, 984}

\bibitem[\protect\citeauthoryear{{Yuan}, {Wu}  \& {Bu}}{{Yuan}
  et~al.}{2012a}]{Yuan2012}
{Yuan} F.,  {Wu} M.,   {Bu} D.,  2012a, \mn@doi [\apj]
  {10.1088/0004-637X/761/2/129}, \href
  {http://adsabs.harvard.edu/abs/2012ApJ...761..129Y} {761, 129}

\bibitem[\protect\citeauthoryear{{Yuan}, {Bu}  \& {Wu}}{{Yuan}
  et~al.}{2012b}]{Yuan2012b}
{Yuan} F.,  {Bu} D.,   {Wu} M.,  2012b, \mn@doi [\apj]
  {10.1088/0004-637X/761/2/130}, \href
  {http://adsabs.harvard.edu/abs/2012ApJ...761..130Y} {761, 130}

\makeatother
\end{thebibliography}

\bsp	
\label{lastpage}
\end{document}